# New Observations of the Eclipsing Binary System NY Vir and its Candidate Circumbinary Planets

Huseyin Er,[1] Aykut Özdönmez,[1] Ilham Nasiroglu [1]★
[1] *Ataturk University, Departments of Astronomy and Space Science, Yakutiye, 25240, Erzurum, Turkey*

**ABSTRACT**

As a result of various studies, it has been determined that several post-common envelope eclipsing binaries have variations in their orbital periods. These variations are thought to be caused by the existence of additional bodies in the system (hypothetical stars or planets) and/or other physical effects (such as angular momentum loss, magnetic activity) of the binary system. It is also known that the sdB+M eclipsing system NY Vir has shown such variations in the last decade, indicating additional objects and/or other physical effects. In this work, we present 51 new eclipse times for this system, which extend the time span of it is $O - C$ diagram by about three years, obtained between 2015 and 2021 using two different telescopes in Turkey. The data obtained in the last 3 years shows a new trend in the $O - C$ diagram differently from the predictions of the previous studies. Our model is consistent with the new $O - C$ diagram, which is statistically well fitted with the quadratic term and the additional two planets with masses of $M_3 = 2.74\ M_{\rm Jup}$ and $M_4 = 5.59\ M_{\rm Jup}$. However, the orbital period variation can also be related to magnetic activity. In order to better understand the mechanism causing the changes in the orbital period, new observation data is needed that will show at least one full cycle of the change in the O-C diagram.

## 1 INTRODUCTION

Three decades have passed since the discovery of the first planet orbiting other stars (Wolszczan & Frail 1992). More than 4500 exoplanets have been discovered to date, thanks to improved instruments on the ground and in space, and more than 5000 additional candidate exoplanets wait for confirmation. A very large part of these planets has been discovered with the so-called Transit Method[1] (http://exoplanet.eu). One of the most impressive discoveries with the Kepler satellite using the transit method is the discovery of a circumbinary planet (CBP) that orbits both binary components (also referred to as P-type orbits by Dvorak (1984) of the Kepler 16 (AB; (Doyle et al. 2011)). Until the decisive discovery of Kepler-16 b, the presence of CBPs remained doubtful (Doyle & Deeg 2018). However, only a small number of these planets were found to exist around binary stars, in spite of more than half of all stars being in multi-star systems (Martin 2018). CBPs formation should be a common event and therefore it is expected that at least several million of them should exist in our galaxy (Welsh et al. 2012). Consequently, studying such planets can lead to a better understanding of the formation and evolution of planets and their host binary stars, which can be rather different from the case of single stars (Lee et al. 2009).

The idea of planets orbiting the binary systems was reported before the Kepler discoveries for eclipsing binaries CM Dra (Deeg et al. 2008) and HW Vir (Lee et al. 2009) by using the timing method (Irwin 1952), which is the same method used in the first exoplanet discovery (Wolszczan & Frail 1992). This method is the most remarkable planet detection method applied for eclipsing binaries such as Post-Common Envelope Binary stars (PCEBs) and Cataclysmic Variables (e.g. Goździewski et al. 2012, 2015; Horner et al. 2012; Marsh et al. 2014; Nasiroglu et al. 2017). Compared to other methods, this method is very sensitive for low-mass binaries and/or massive planets with long orbital periods (Sterken 2005; Pribulla et al. 2012). The timing method is based on the cyclic variations in the $O-C$ (Observed – Calculated) times of the stellar eclipses, which is interpreted as the Light Travel Time (LTT) effect phenomenon. These variations can result from the gravitational effects of a distant orbiting third body (e.g., a CBP) which leads to the binary system rotating around the barycenter of this triple system, causing the timing of the eclipses to appear slightly early, on time, or late, etc. in a periodic manner. This LTT effect can be determined and used to investigate the existence of planetary-mass companions in such systems. However, it should be cautious to certainly claim the CBPs until there is independent evidence on their existence (Marsh 2018). Because there are other mechanisms that can cause structural changes in such systems i.e. that will result in orbital period change; e.g. the Applegate mechanism (Applegate 1992).

NY Vir (PG 1336-018) was discovered through the Palomar-Green Survey (Green, Schmidt, & Liebert 1986). Later, Kilkenny et al. (1998) identified this system as an HW Vir-type eclipsing binary with an orbital period of about 0.1010174 days, contains a very hot sub-dwarf B (sdB) which is also a rapid pulsator and a late dwarf M5-type companion. Kilkenny et al. (1998) also presented the effective temperatures of $T_1 = 33000K$ and $T_2 = 3000K$, the relative radii of $r_1 = 0.19$ and $r_2 = 0.205$, an orbital inclination of 81°, and the mass of the primary as 0.5 $M_\odot$ leads to a mass ratio $q = 0.3$ for the system. A detailed photometric and spectroscopic investigation of the pulsation properties of the sdB component can be found in (Kilkenny et al. 1998, 2000; Hu et al. 2007; Vučković et al. 2007, 2009; Van Grootel et al. 2013).

---

★ E-mail: inasir@atauni.edu.tr
[1] http://exoplanet.eu/

After its discovery, the eclipse timing variation of NY Vir has studied by many authors, and they claimed the period changes for this system (a detailed list of the eclipse times can be found in Table 1). The first significant variation in the orbital period was reported by Kilkenny (2011) with a decrease rate of $-11.2 \times 10^{-13}$ d per orbit, revealed a strong quadratic effect. Then as well, Çamurdan et al. (2012) reported also a decreasing orbital period ($dP/dt = -4.09 \times 10^{-8}$ day yr$^{-1}$), using the quadratic term, indicating an angular momentum loss of the binary star system. Soon after, Qian et al. (2012) detected a small-amplitude cyclic variation with a period of 7.9 yr and a semi-amplitude of 6.1 s by adding a sinusoidal term to the quadratic model, ascribed to the presence of a third body with masses of $M_3 \sin i' = 2.3(\pm 0.3) M_{Jup}$. Later, Lee et al. (2014) discussed the 2-planet model by using their 39 new eclipse times, with periods of $P_3 = 8.2$ yr and $P_4 = 27.0$ yr, semi-amplitudes of $K_3 = 6.9$ s and $K_4 = 27.3$ s, and minimum masses of $M_3 \sin i_3 = 2.8\ M_{Jup}$ and $M_4 \sin i_4 = 4.5\ M_{Jup}$. The models so far have been found to be incompatible with the new observations. Recently, Pulley et al. (2018) published new eclipse times that show a deviation from the predictions of the previous studies, in the $O - C$ diagram of NY Vir. Finally, Song et al. (2019) published 18 new eclipse times between 2018 January 24 and April 18 for this system. They reported a quadratic term with a rate of $dP/dt = 2.83 \pm 0.28 \times 10^{-12}$ s s$^{-1}$ using the two-planet solution and suggested the presence of two planets with orbital periods of $8.64 \pm 0.17$ yr and $24.09 \pm 0.65$ yr, and minimum masses of $2.66 \pm 0.36\ M_{Jup}$ and $5.54 \pm 0.28\ M_{Jup}$.

In this study, we present 51 new mid-eclipse times of NY Vir, that extend the time span of the $O - C$ diagram by about three years, obtained between 2015 March 25 and 2021 March 13. The span of the $O - C$ diagram of this system covers 25 years along with the published eclipse times in the literature and our new data. In Section 2, we present the observations and data reduction process, as well as determination of the eclipse times. Section 3 presents analysing on $O - C$ diagrams and system parameters of NY Vir. In Sections 4, we discuss and conclude our results.

## 2 OBSERVATIONS AND DATA REDUCTION

We present new photometric observations of NY Vir performed between 2015 March 25 and 2021 March 13 with 0.6 m telescope equipped with 1k x 1k Andor iKon-M934 CCD camera (with the field of view of 11.4′x 11.4′and pixel scale of 0.67"/pixel) at the Adiyaman University Observatory (ADYU60, Adiyaman, Turkey) and 1 m telescope equipped with a 4k × 4k SI1100 CCD camera, the field of view of 21.5′x 21.5′and pixel scale of 0.31"/pixel at the TÜBİTAK National Observatory (TUG T100, Antalya, Turkey). We performed the observations using no filter for an exposure time in the range of 10-15 seconds for T100 (readout time: ~ 14 s) and 10-25 seconds for ADYU60 telescopes (readout time: ~ 1 s) to have enough counts and proper time resolution for the light curves of NY Vir. Besides, high SNR in range of ~ 80 − 800 based on the weather condition was obtained during the observations. The typical value of the seeing during the entire observation period was in the range of 0.8" - 1.1" for T100 and 1.7"-3.1" for ADYU60. These seeing and SNR values show that the observations made in this study are of relatively high quality (especially T100).

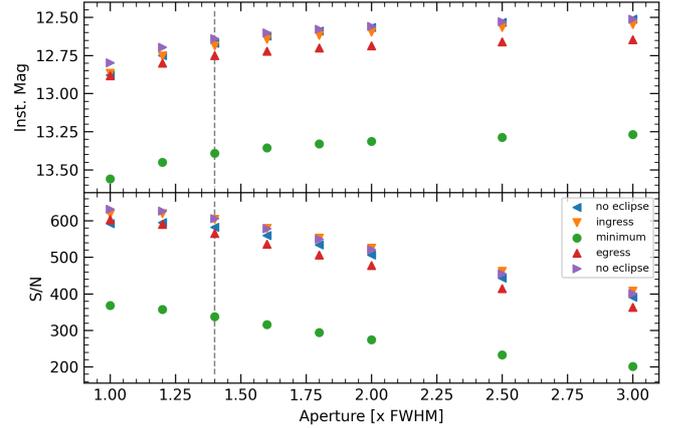

**Figure 1.** The instrumental magnitude and S/N distribution regarding multiple aperture radii. The different colors stand for approximately different part of the light curve on 2020-03-12 obtained with the TUG T100 telescope. The vertical dashed line represent aperture at $1.4 \times$ FWHM.

The standard CCD reduction was performed on the data obtained from the observations, i.e. the bias subtraction, flat-field and cosmic-ray correction were applied to all scientific images. Because the dark counts were negligible for CCDs attached to the used telescopes, dark subtraction was not done. After CCD reduction, we used a developed script using Python and Sextractor to implement aperture photometry. For the differential photometry, we chose one comparison star which is nearby and has similar brightness with the source. However, we also compared the comparison star with another star to see if there is any flux variability. In order to determine the optimal aperture radius, we select 8 different aperture radii from 1 × FWHM to 3 × FWHM. The instrument magnitude and signal-to-noise ratio (S/N) regarding aperture radii for a different part of the light curve on 2020-03-12, i.e. no eclipse, ingress, egress, minimum, are represented in Fig. 1 which shows an increasing trend in instrument magnitude on the contrary to the trend in the S/N with increasing aperture radii. The instrument magnitudes are roughly invariable using apertures from 1.8 × FWHM to 3. × FWHM. Since all trends on approximately different parts of the light curve as well as on different nights are similar, the aperture radius of 1.4 × FWHM is adopted, when the magnitudes are near the average magnitude of the whole data. As well as, we got considerably high (~ 350) SNR values for this aperture. Therefore, we determined differential magnitudes, and then obtained the 51 eclipse light curves of NY Vir.

The light curves of NY Vir were modelled with a modified gaussian profile (hereafter it will be called as Gauss*Poly) which is formed by multiplying inverse Gauss function $G(\tau)$ with a polynomial function $P(\tau)$, as in Beuermann et al. (2012). The eclipse times were derived from the 51 modelled light curves as given in Fig. 2 with their residuals. The eclipse times were converted to the barycentric dynamical julian time (BJD) from julian date (JD), using the procedure in Eastman, Siverd, & Gaudi (2010). The derived and collected eclipse times from the literature are listed in Table 1. We also calculated the root mean squares (rms) for the fitting results. The highest rms value is 0.057 mag, while the mean rms value is 0.024 mag. The

**Table 1.** The Mid-eclipse Times of NY Vir, its error and references.

| In literature | | |
|---|---|---|
| BJD | error | References |
| 2450223.36202800 | 0.000040 | Kilkenny et al. (1998) |
| 2450223.46315800 | 0.000020 | Kilkenny et al. (1998) |
| 2450224.37222800 | 0.000030 | Kilkenny et al. (1998) |
| ... | ... | ... |
| From our data | | |
| BJD | error | References |
| 2457106.59016795 | 0.000020 | This work (TUG T100) |
| 2457164.27030104 | 0.000039 | This work (ADYU60) |
| 2457166.29063049 | 0.000025 | This work (TUG T100) |
| 2457166.39160512 | 0.000032 | This work (TUG T100) |
| ... | ... | ... |

* Full table is available in its entirety in machine-readable form.

rms values demonstrate that our results for fitting the observational light-curves are reliable.

On the other hand, the pulsation of the sdB component is clearly visible in the residuals of some light curves. For quick measurement, we used Lomb-Scargle analysis on residuals of 17 best quality light curves (i.e. rms≤ 0.015 mag), and found moderate signals that indicate a pulsation period between 170 and 186 s, but 4 of them show very weak signals placed in the false alarm limit. A detailed light curve analysis by Vučković et al. (2007) shows that the amplitudes of the pulsation are weaker during the primary eclipse of the system. Since we do not know the amount of time shift due to pulsation, we had to assume that this pulsation effect on mid-eclipse times is negligible during our analysis.

## 3 LTT MODEL OF THE ORBITAL PERIOD VARIATION

We combined our present observations with previously published eclipse times in Kilkenny et al. (1998, 2000); Kilkenny (2011, 2014), Çamurdan et al. (2012), Qian et al. (2012), Lee et al. (2014), Lohr et al. (2014), Pulley et al. (2018), Baştürk & Esmer (2018), Song et al. (2019). All eclipse times are listed in Table 1. We first made a few amendments to these eclipse times in a similar way to Song et al. (2019); i.e. (i) For the uncertainties of the eclipse times in Kilkenny et al. (2000), since there is no information about its errors, we assumed 0.00005 days for all uncertainties. (ii) Since there is a few outliers in Çamurdan et al. (2012); Qian et al. (2012); Lee et al. (2014); Baştürk & Esmer (2018), which differs more than three standard deviations from the overall (O-C) trend, even if the eclipse times were obtained on similar dates, we removed these 31 outliers during LTT modelling. Even so we added all this literature data to our Table 1, and we labeled which mid-eclipse times were used or not. (iii) We added 145 eclipse times of WASP observations given by Lohr et al. (2014) to our Table 1 (in total 342 mid-eclipse times). However, we binned these data in four groups based on their times, and only used these four binned mid-eclipse times of WASP observations during modelling. (iv) We added the eclipse times calculated from the American Association of Variable Star Observers (AAVSO)[2] in Song et al. (2019). As well as we presented 51 new mid-eclipse times of NY Vir obtained from our observations between 2015 March 25 and 2021 March 13. Finally, to examine the $O-C$ diagram, we used 166 (including 4 binned WASP data) of 197 mid-eclipse times obtained for NY Vir (from 1996 to 2021) after a few modifications for literature data.

Using the eclipse times in Table 1, we derived initial ephemeris time and orbital period of the binary system from the following linear ephemeris equations by minimizing chi-square using Python library of *LMFIT* (Newville et al. 2014).

$$T_{eph}(E) = T_0 + E \times P_{bin}$$
$$= \text{BJD } 2453174.442636(12) \quad (1)$$
$$+ E \times 0.1010159686(3)$$

Here, linear ephemeris $T_{eph}$ is the mid-eclipse time in BJD. $T_0$ is the initial ephemeris for the mid-eclipse time at zero cycle ($L = 0$), and $P_{bin}$ is the orbital period of the binary system. We obtained the $O-C$ times as shown in Fig. 3.

As seen in Fig. 3, there is a cyclic variation in the $O-C$ diagram that can be explained with the existence of the gravitational effect perturbing the center of mass of the binary system. In literature, Qian et al. (2012) explained this variation with a quadratic ephemeris and one planet, and Lee et al. (2014) used a model contains linear ephemeris with two planets. Recently, Song et al. (2019) added newest observation and modelled the variation of the $O-C$ diagram with quadratic ephemeris and two planets. We tested all models with their parameters in the literature to explain this variation (see Fig. 3), but these models are not consistent with recent $O-C$ data. Thus, we used the following model of the quadratic ephemeris with two planets to update the system parameters (Song et al. 2019).

$$O-C \text{ [days]} = \beta E^2 + \tau_1(E) + \tau_2(E) \quad (2)$$

Here, $\beta$ is the quadratic term given by $P\dot{P}/2$, where $\dot{P}$ is derivation of the period concerning to ephemeris time (i.e $dP/dt$). The following expression is for the light travel time (LTT) term of $\tau_{1,2}$ of the 1st and 2nd planets (Irwin 1952);

$$\tau_{1,2} = K_i \frac{1}{\sqrt{1-e_i^2 \cos^2 \omega_i}} \left[ \frac{1-e_i^2}{1+e_i \cos \nu_i} \sin(\nu_i + \omega_i) + e_i \sin(\omega_i) \right] \quad (3)$$

where $e_i$ and $\omega_i$ is the eccentricity and the longitude of pericenter of the $i$th planet's orbit (i=3 for the first planet, i.e. 3rd body in the system), respectively. Another orbital parameter $\nu_i$ is the true anomaly of the binary orbit around the system's barycenter. The orbital period of the $i$th body, $P_i$, and the time of pericenter passage of $t_{0,i}$ are hidden in $\nu_i$ calculation (eq. 11 and 12 in Irwin 1952). To calculate $\nu_i$, we used the iteration process as given in Odell & Gooding (1986, using the starting formula S9 in Table 1). The semi-amplitude of LTT signal in $O-C$ arising by perturbation of $i$th planet, $K_i$, can be calculated as follows;

$$K_i = \frac{a_{12} \sin i_i \sqrt{1-e_i^2 \cos^2 \omega_i}}{c} \quad (4)$$

where $a_{12} \sin i_i$ is the projected semi-major axis of the binary system around the triple system (binary companions and $i$th planet), and $i_i$ is the inclination of the planet's orbit. In order to fit the variation of $O-C$ times, we used the model with eleven free parameters ($dP/dt$, $a_{12} \sin i_{3,4}$, $e_{3,4}$, $\omega_{3,4}$, $t_{03,04}$ and $P_{3,4}$) that contains the quadratic ephemeris and the LTT effects of two planets. In models with many free parameters such as LTT models, determination of parameters by typical methods does not give quality results that take into account parameter degeneration, the convergence of parameters, and

[2] https://www.aavso.org/

4  *H. Er et al.*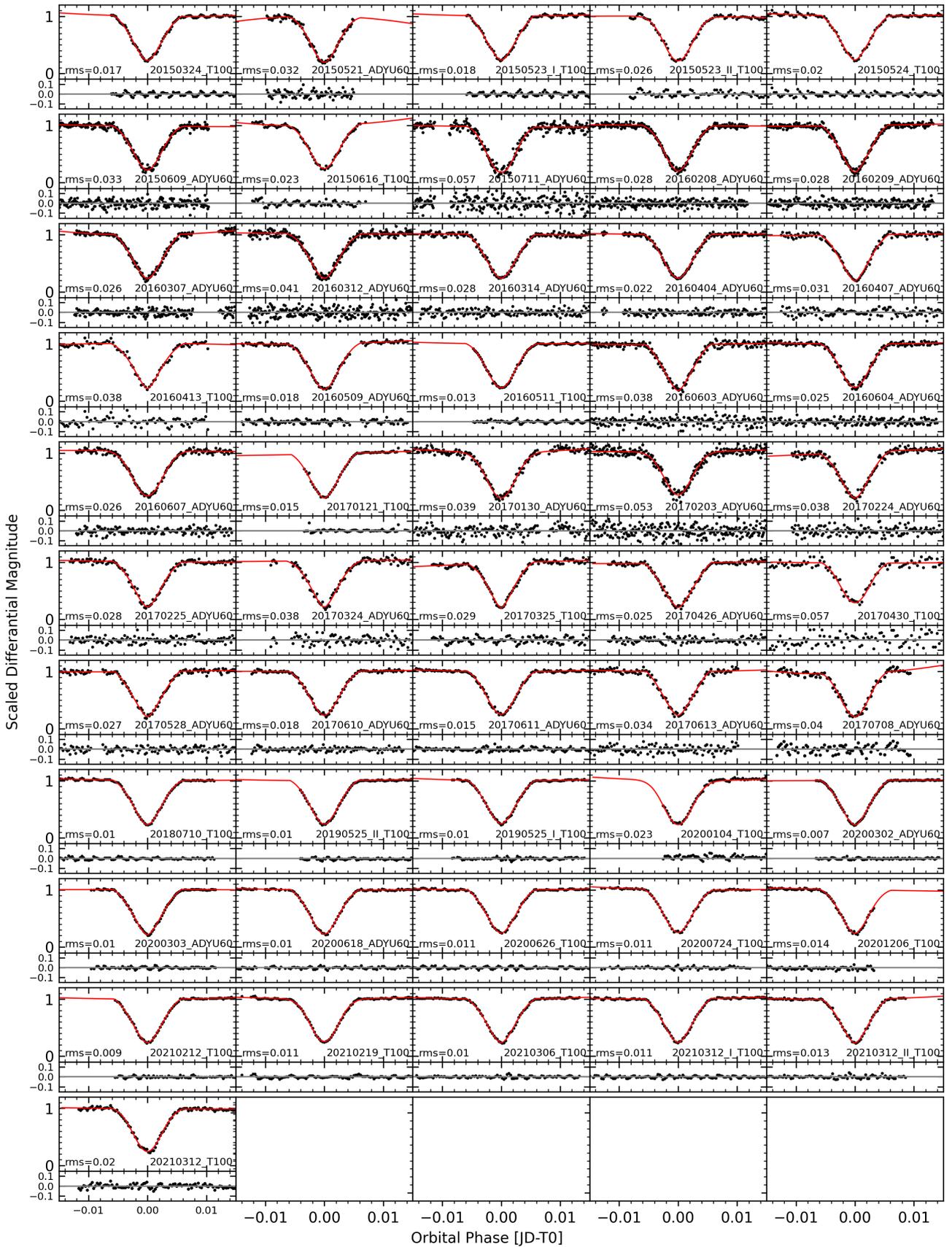

**Figure 2.** Light curves of NY Vir obtained from observations of TUG T100 and ADYU60. Light curves were fitted with Gauss*Poly function as in Section 2. The date of observation and used telescope are labeled. Roman numbers as I and II describe the sequence of total eclipsing during the same day. Calculated rms values in units of magnitude are also denoted.

MNRAS **000**, 1–10 (2021)

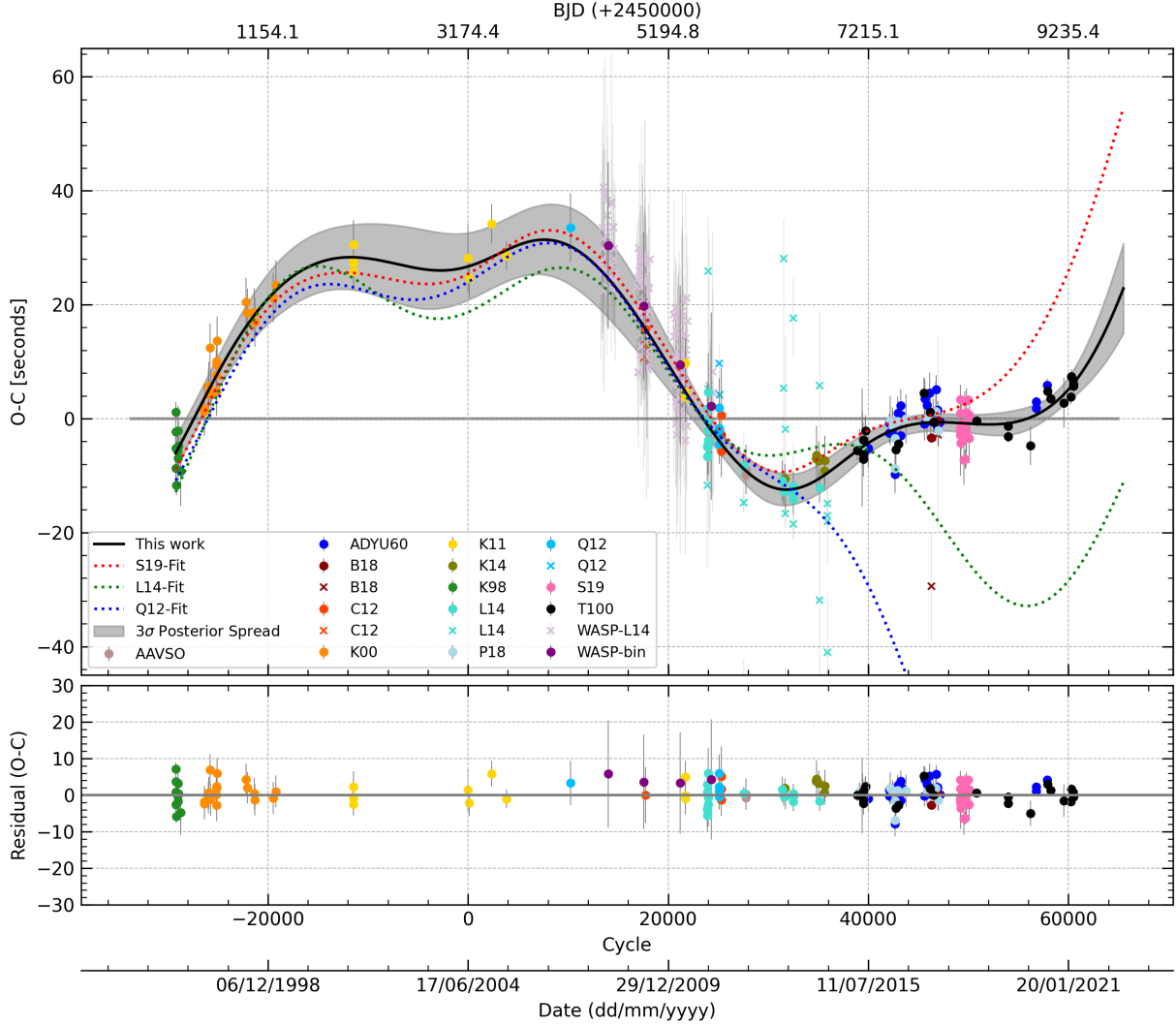

**Figure 3.** Upper plot: $O - C$ Diagram of NY Vir, linear ephemeris times are calculated from Equation 1, which is represented by black filled circles for TUG T100 and blue filled circles for ADYU60, with all data in literature given in different colors which are labelled with the corresponding abbreviations. Here, abbreviations are for references; K98 for Kilkenny et al. (1998), K00 for Kilkenny et al. (2000), K11 for Kilkenny (2011), C12 for Çamurdan et al. (2012), Q12 for Qian et al. (2012), K14 for Kilkenny (2014) WASP-L14 for Lohr et al. (2014), L14 for Lee et al. (2014), B18 for Baştürk & Esmer (2018), P18 for Pulley et al. (2018), and S19 for Song et al. (2019). Note: the eclipse times obtained from AAVSO data are taken from Song et al. (2019). The black line shows the model obtained in our study. The shaded gray area represents the $\pm 3\sigma$ posterior spread which is calculated from 1000 randomly selected parameter samples from the MCMC posterior. We reproduced the other LTT models using the system parameters of NY Vir given in the references in question, e.g. red dotted line for the model of quadratic ephemeris and LTT effects from two planets in Song et al. (2019), green dotted line for the model of two LTT effects in Lee et al. (2014), blue dotted line for the model of quadratic ephemeris and one LTT effect in Qian et al. (2012). Lower plot: The residual of $O - C$ times for the model obtained in this paper. The cycles calculated using Equation 1, i.e. E=0 is at $T_0 = 2453174.442636$. The rms value of the residuals is calculated as 8.4 seconds.

probability distributions. Therefore, the Markov-Chain Monte Carlo (MCMC) method is more convenient to investigate the variation of $O - C$ times (e.g. Goździewski et al. 2015; Nasiroglu et al. 2017).

To express the $O - C$ variability through our model as given in Equation 2, we optimized the likelihood function $\mathcal{L}$.

$$\ln \mathcal{L} = -\frac{1}{2} \sum_i^N \frac{(O-C)_i^2}{\sigma_i^2 + \sigma_f^2} - \sum_i^N \ln \sqrt{\sigma_i^2 + \sigma_f^2} - N \ln \sqrt{2\pi} \quad (5)$$

Here, $(O - C)_i$ denotes the deviation of observed $i$th eclipsing time from calculated from Eq. 1, i.e $(T_{obs}(E_i) - T_{eph}(E_i))$, and $\sigma_i$ is for the measurements uncertainties of observed $i$th eclipsing time, where $i = 1, 2, 3, ..., N$ (Goździewski et al. 2015). To this point, systematic uncertainties have not been considered during obtaining uncertainties of mid-eclipse times, but systematic uncertainties should be taken into account during the fitting process. Thus, a free parameter $\sigma_f$ was included in the likelihood function to fit for the fractional amount by which the uncertainties are underestimated (i.e. due to $\sigma_i$ do not contain systematic uncertainties). $\sigma_f$ has units of the day as similar with $\sigma_i$, and it scales the raw uncertainties $\sigma_i$ in quadrature.

The general goal of optimization of the dynamical model through the MCMC algorithm is investigating the space of the free model parameters from the given number of samples from the posterior probability distribution. To sample the posteriors, we use the Markov Chain Monte Carlo (MCMC) *emcee* package of the affine-invariant

**Table 2.** Prior Parameters obtained from GA

| Parameters | Units | Ranges | Values |
|---|---|---|---|
| $dP/dt$ | s s$^{-1}$ | $[10^{-14}, 10^{-11}]$ | $2.33 \times 10^{-12}$ |
| $a_{12} \sin i_3$ | au | $[0, 0.1]^*$ | 0.0188 |
| $a_{12} \sin i_4$ | au | $[0, 0.1]^*$ | 0.0631 |
| $e_3$ | | $[0, 0.9]$ | 0.077 |
| $e_4$ | | $[0, 0.9]$ | 0.094 |
| $\omega_3$ | deg | $[0, 360]$ | 318 |
| $\omega_4$ | deg | $[0, 360]$ | 347 |
| $t_{0,3}$ | BJD | $\geq 2449000$ | 2456488 |
| $t_{0,4}$ | BJD | $\geq 2449000$ | 2450453 |
| $P_3$ | yr | $\geq 0$ | 8.93 |
| $P_4$ | yr | $\geq 0$ | 26.03 |
| $\sigma_f$ | s | $\geq 0$ | 1.64 |

* The maximum value calculated by assuming $K_{3,4} \leq 50$ sec (see Fig. 3)

ensemble sampler (Goodman & Weare 2010), kindly provided by (Foreman-Mackey et al. 2013).

MCMC strongly depends on the prior parameters, and so we used the Genetic Algorithm (GA) similar methodology in Charbonneau (1995) to determine the prior parameters of MCMC. The ranges of the free parameters imposed while running the GA and the priors obtained from the GA are listed in Table 2. Using these priors, we ran the MCMC to sample the posteriors for 512 initial conditions (i.e. walker) and tested these walkers in 100,000 samples (i.e. steps in chains of each free parameter). To set the number of samples, we derived auto-correlation time convergence with the chain number. For the acceptance fraction of 0.5 which indicating a confident estimate from the MCMC sampler, the shorter samples (i.e. $\lesssim 50,000$) may give relatively unreliable results while the longer samples (i.e. $\gtrsim 100,000$) may be considered redundant (for details see Foreman-Mackey et al. 2013). Thus we obtained the most plausible system parameters through maximizing $\mathcal{L}$ as given in Table 3.

The best-fitting parameters of the model is shown in the $O-C$ diagram in Fig. 3. The 1-D and 2-D posterior probability distributions of the system parameters sampled by MCMC are illustrated in the corner plot (Fig. 4).

The additional planet parameters can also be calculated using the related parameters in Table 3. In practically, the companion masses can be determined from the mass function given below;

$$f(M_i) = \frac{(M_i \sin i_i)^3}{M_{tot}^2} = \frac{4\pi^2 (a_{12} \sin i_i)^3}{G P_i^2} \quad (6)$$

It should be noted that only minimum mass could be calculated if the inclination is not known as in our study. Hence, adopting the total mass of the system as $M_{tot} = \sim 0.59\,M_\odot$ (Model II in the Table 3 in Vučković et al. 2007), the lower limit of masses of the planets are calculated as $M_3 = 2.74 M_{\rm Jup}$ and $M_4 = 5.59 M_{\rm Jup}$.

## 4 DISCUSSION AND CONCLUSIONS

We present 51 light curves of NY Vir observed between 2015-2021 with ADYU60 and TUG T100 telescopes in Turkey. By fitting these light curves with Gauss*Poly model, we obtained the latest eclipse times for the $O-C$ diagram of NY Vir. We should note that Gauss*Poly model is more proper for the light curve of NY Vir since a few light curves have a sloping upward trend at the outside of the eclipse. Polynomial function in the model represents the outside of the eclipse while the gaussian function is only for the eclipse between ingress and egress (Beuermann et al. 2012). Regarding the residuals of light curves, we obtained moderate signals by the quick Lomb-Scargle analysis that lead us to estimate the pulsation period in the range of 170-186 s. However, we neglected the possible variation effect on mid-eclipse time due to stellar pulsation since it is beyond the scope of this study. This assumption is reasonable considering the results of Vučković et al. (2007) in which the amplitude of pulsation is found smaller during the primary eclipses.

Combining our eclipse times with the ones obtained in the literature, we listed a total of 342 mid-eclipse times (145 of them is WASP data) in Table 1 from 1996 to 2021, 166 (including 4 binned WASP data) of them are available for the analysis after a few modifications. Using these eclipse times, we obtained the best-fitting parameters of linear ephemeris, i.e. initial ephemeris at $E = 0$ is $T_0 = $ BJD 2453174.442636 and orbital period of binary is $P_{bin} = 0.10101596862$ days. We formed the $O-C$ diagram using mid-eclipse times collected from our observations and literature (Fig. 3). The $O-C$ diagram spans $\sim 25$ years, based on the literature and six years of our new data that extend the time span of the $O-C$ diagram by about three years.

Before commenting on the $O-C$ model, we would like to discuss the eclipse times (or rather $O-C$ times). In this study, we present all light curves with their models and rms values to show how well determined the eclipse times from the light curves. The average rms value is 0.024 mag, while the highest rms is 0.057 mag. Thus we could claim that our eclipse times (and light-curve models) are reliable. However, there is not much information about light curves in most of the previous studies for NY Vir i.e. discussion on scattering or fit statistics or all light-curves are not given (only a few examples) despite pulsation. Thus, we can not be sure about the precision of the eclipse time measurements. This uncertainty can be also seen in the $O-C$ diagram (Fig. 3) owing to outliers scattered from the general trend of $O-C$ times. Hence, we removed a few outliers that differ more than three standard deviations from the overall trend during LTT modelling. We suggest that these outliers exhibit large uncertainties and they strongly scatter from the LTT model shown in this study.

Since our new data extend the archived list of eclipsing times, the previous LTT models explaining the $O-C$ behavior of this system (e.g. Qian et al. 2012; Lee et al. 2014; Song et al. 2019) are not consistent with recent $O-C$ data (Fig. 3). Thus, our new observations make it possible to improve the system parameters and put constraints on previous LTT models. The earliest studies consisted of the $O-C$ data less than half cycle of a quasi-sinusoidal modulation in which the orbital period variation of NY Vir had a downward slope until 2013, and they used models of the one planet with or without quadratic effect (Kilkenny 2011; Qian et al. 2012; Çamurdan et al. 2012). It is obvious that the variation in the $O-C$ diagram could be explained by the model of quadratic ephemeris with LTT effects from potential two planets as in Song et al. (2019). In the light of the above conclusion, we fitted the model of quadratic ephemeris with two LTT effects to explain the $O-C$ variation by running GA and MCMC algorithms, and so we determined the system and orbital parameters of NY Vir as listed in Table 3. These best-fitting parameters from MCMC (Table 3) are close to the initial values obtained from the GA (Table 2).

The most noticeable result of the MCMC rather than the least-square method is 1D and 2D posteriors distributions of the free parameters in the model as shown in Fig. 4. This is useful because it demonstrates all of the covariances between parameters. All the 1D distributions show single strong peaks while samples have relatively uniform 2-D distributions around a single solution Fig. (4). Besides, we could demonstrate that there is no bias or degeneration in the parameter space. However, the solutions on some parameters scatter in large ranges, i.e. $e_3$ from 0.00 to 0.24, $\omega_3$ from 286° to 461°, the differences between an upper and lower limit of $t_{0,3,4}$ are ∼ 1600 and ∼ 2000 days, respectively, under consideration of 1$\sigma$ limits. These uncertainties arise from especially lack of observation before 2013, it can also be seen in ±3$\sigma$ posterior spread which was calculated from 1000 randomly selected models from MCMC samples (Fig. 3). To reduce the uncertainties, a full cycle involving further observations is needed. However, almost all $O - C$ data with considering their errors are placed in the ±3$\sigma$ posterior spread, except outliers.

**Table 3.** System Parameters for NY Vir.

| Parameters | Unit | Lee et al. (2014) | Song et al. (2019) | This Work |
|---|---|---|---|---|
| Binary Star System | | | | |
| $T_0$ | BJD | 2453174.442699(91) | 2453174.442647(13) | 2453174.442636(12) |
| $P_0$ | day | 0.1010159668(43) | 0.1010159677(4) | 0.1010159686(3) |
| $dP/dt$ | s s$^{-1}$ | – | $2.83(0.25) \times 10^{-12}$ | $2.64^{+0.21}_{-0.24} \times 10^{-12}$ |
| First Planets | | | | |
| $a_{12} \sin i_3$ | au | $0.0153 \pm 0.0220$ | $0.0151 \pm 0.0082^*$ | $0.0161^{+0.0018}_{-0.0016}$ |
| $e_3$ | | | $0.15 \pm 0.08$ | $0.12^{+0.12}_{-0.12}$ |
| $\omega_3$ | deg | $346 \pm 15$ | $348 \pm 6$ | $351^{+69}_{-52}$ |
| $t_{0,3}$ | BJD | 2453472(141) | 2453472 | $2456762^{+628}_{-479}$ |
| $P_3$ | yr | $8.18 \pm 0.18$ | $8.64 \pm 0.17$ | $8.97^{+0.36}_{-0.24}$ |
| $K_3$ | s | $6.9 \pm 1.0$ | $7.6 \pm 0.7$ | $7.9^{+1.0}_{-0.8}$ |
| $M_3 \sin i_3$ | $M_{\rm Jup}$ | $2.78 \pm 0.19$ | $2.66 \pm 0.26$ | $2.74^{+0.37}_{-0.34}$ |
| Second Planet | | | | |
| $a_{12} \sin i_4$ | au | $0.0550 \pm 0.0160$ | $0.0625 \pm 0.0047^*$ | $0.0686^{+0.0041}_{-0.0039}$ |
| $e_4$ | | $0.44 \pm 0.17$ | $0.15 \pm 0.01$ | $0.19^{+0.09}_{-0.07}$ |
| $\omega_4$ | deg | $333 \pm 15$ | $320 \pm 4$ | $398^{+42}_{-34}$ |
| $t_{0,4}$ | BJD | 2450031(497) | 2450031 | $2451694^{+1108}_{-787}$ |
| $P_4$ | yr | $27.0 \pm 0.37$ | $24.09 \pm 0.65$ | $27.2^{+1.3}_{-1.2}$ |
| $K_4$ | s | $27.3 \pm 8.2$ | $31.4 \pm 1.1$ | $33.8^{+2.4}_{-2.9}$ |
| $M_4 \sin i_4$ | $M_{\rm Jup}$ | $4.49 \pm 0.72$ | $5.54 \pm 0.20$ | $5.59^{+0.51}_{-0.49}$ |

$^*$ $a_{12} \sin i_{3,4}$ calculated by us from best-fit parameters of Song et al. (2019)

We also calculated root-mean-square (rms) of the residual of $O - C$ times as 8.4 s. Since the $O - C$ residual for the similar dates are scattering up to 8 s, the rms value are consistent with uncertainties of the observational data (see residual $O - C$ diagram in Fig. 3). In addition, the MCMC posterior sampling reveals relatively low $\sigma_f = 1.7 \pm 0.2$ s for the systematic error floor of the observations considering the model, which is similar to mean observational uncertainties of 2.8 s. Without fractional correction, the raw reduced $\chi^2_\nu \sim 1.7$ will be scaled underestimated uncertainties. This systematic error should be considered in further studies in which mid-eclipse times of NY Vir are used.

All of the results demonstrate that our model is statistically well consistent with current observations. However, it should be noted that the rms scatter and mean total error (observational and systematic, 4.5 s) in the O-C is quite similar to the orbit light-travel-time delay of $K_3 = 7.9$ seconds. This raises a faint suspicion on the significance of two planet models. Later in this paper, we suggest another mechanism may also exist to explain O-C times of NY Vir.

We also compared our results with the studies in the literature which considered two planets model even though our recent observations are not consistent with prediction of these studies (see Fig. 3). Almost all of the most plausible system parameters are relatively in agreement with previously obtained in the literature within errors (see Table 3). It is expected that the system parameters (especially $\omega_4$, $t_{0,4}$, $P_4$) for the second planet should be updated, looking at the trend of $O - C$ since rising toward the second cycle that begins after March 2020 even though Song et al. (2019) predicted that should begin in 2019. On the other hand, Lee et al. (2014) didn't account for the quadratic effect in their two planets models. Song et al. (2019) used different fixed planet's orbital eccentricities ($e_3$, $e_4$) and adopted the time of pericenter passage of planets ($t_{0,3}$, $t_{0,4}$) that are given in Lee et al. (2014). However, we set free all eleven system parameters in our model. Thus, comparing eccentricities and times of pericenter passage, which were obtained in literature and in our work, will be unsatisfactory. We also comment on our larger uncertainties than previously obtained in the literature (see Table 3). Since the MCMC method searches a larger number of the free parameters in larger ranges using an overwhelming number of samples rather than the least-square method, the larger uncertainties could be expected that represent more reliable results. Again, more observations consisting of a full cycle of O-C times are needed to reduce the uncertainties and prove the model for hypothetical planets even though the LTT model statistically is consistent with the data. Since the model parameters are changing with observations over time, we believe that larger uncertainties should exist.

We determined the projected semi-major axis of the binary star $a_{12} \sin i_{3,4} = 0.016$ and 0.07 au indicating largeness of amplitude of LTT signal, while the eccentricity ($e_3 = 0.12$ and $e_4 = 0.19$) and the longitude ($\omega_3 = 351°$ and $\omega_3 = 398°$) of the pericenter are related to the skewness of LTT signal. These values suggest a small ($K_3 = 7.9$ s) and large ($K_4 = 33.8$) semi-amplitude of the LTT signals arising from the first and second candidate planets, respectively, which also leads us to the heaviness of masses i.e. min $M_3 = 2.74$ $M_{\rm Jup}$ and $M_4 = 5.59$ $M_{\rm Jup}$, respectively. These outputs are in agreement with the previous studies (see Lee et al. 2014; Song et al. 2019). However, it seems that relatively high mass planets are fairly close to each other (∼ 4 au), and this proximity may cause additional orbital variations as well. It should be considered in future studies.

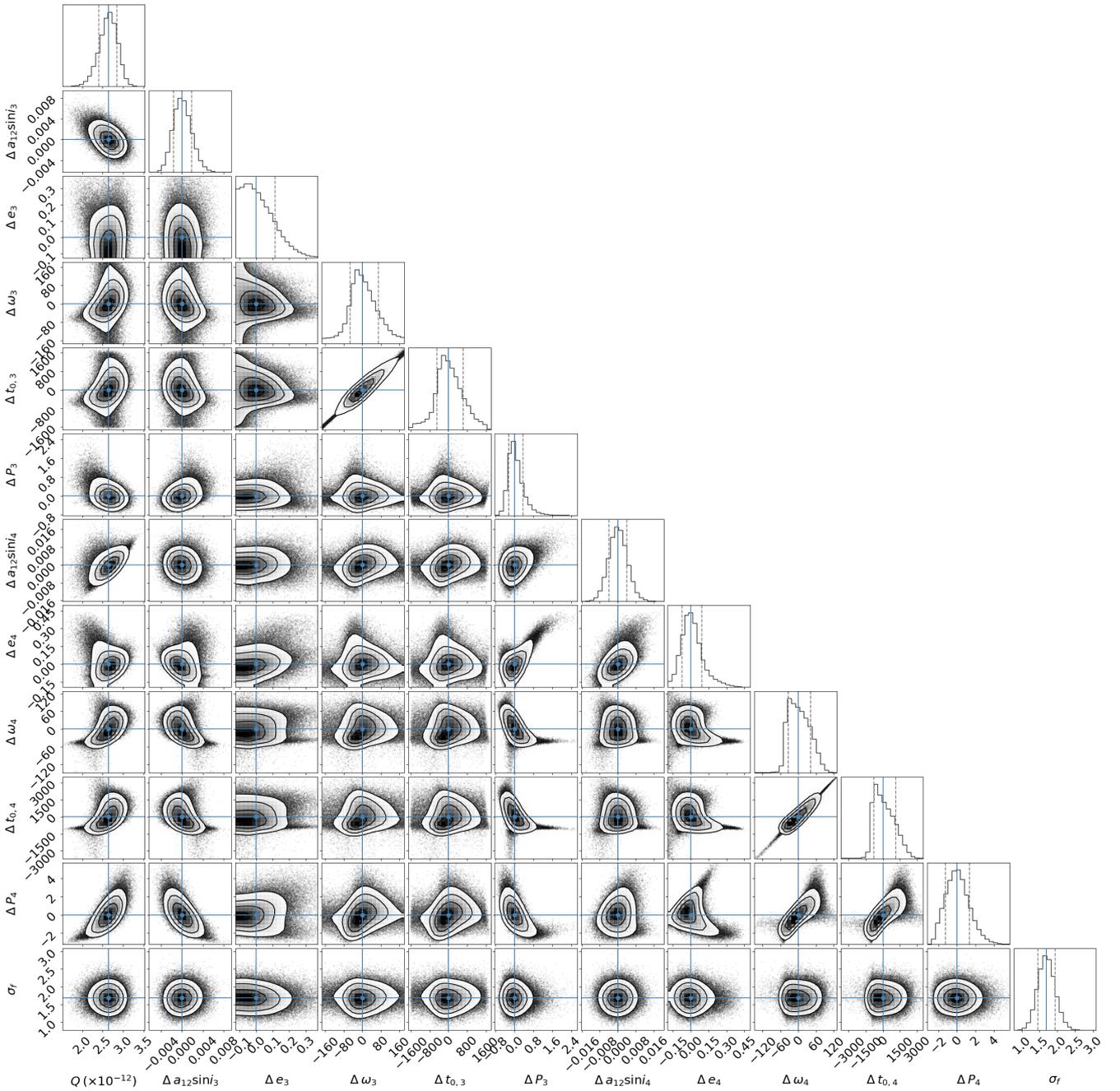

**Figure 4.** 1-D and 2-D projections of the posterior probability distributions of the free parameters inferred from the $O-C$ model. We removed about of the first 100 initial, burn-out samples and thinned the samples by taking one on every 100 samples. $\Delta$ represent differences between posterior and calculated values of the parameter. Contours are for the 16th, 50th, and 84th percentile of samples in the posterior distribution. This figure is made using corner.py (Foreman-Mackey et al. 2013).

The orbital period variation in binary star systems could also be originated in the factors other than N-bodies such as angular momentum gain/loss (Andronov, Pinsonneault, & Sills 2003; Schwarz et al. 2009), magnetic activity (Applegate 1992; Lanza, Rodono, & Rosner 1998) and mass transfer (Rovithis-Livaniou 2006). In the case of NY Vir, the orbital period variation was explained with the angular momentum loss by Çamurdan et al. (2012). However, Qian et al. (2012) and Lee et al. (2014) reported that this variation could not be explained by the loss of angular momentum caused by gravitational radiation and/or magnetic braking, instead, this variation may be due to one and two planetary effects, respectively. Recently, Song et al. (2019) suggested that the orbital period variation is due to the angular momentum gain (i.e. dP/dt is positive) plus the existence of the third and/or fourth bodies orbiting the system.

The model we applied in this study is also well consistent with the O-C diagram which explains the orbital periods variation arising from angular momentum gain ($dP/dt = 2.64 \times 10^{-12}$ s s$^{-1}$) plus two planetary effects. After a while, if the period variation changes its trend from increase to decrease (i.e. dP/dt from positive to negative), it may reveal the presence of a magnetic cycle in the system. We considered the Applegate mechanism in which magnetic activity of the less massive component in a binary system causes this cyclic period variation (Applegate 1992; Lanza, Rodono, & Rosner 1998). According to Applegate's model, the angular momentum transfer could occur under the condition of $\Delta E_{min} \ll E_{sec}$ (Applegate 1992; Tian, Xiang, & Tao 2009). Völschow et al. (2016) investigated Applegete's model for NY Vir, and they reported that magnetic activity insufficient to explain the orbital period variation since the energy discrepancy of $E_{min}/E_{sec} = 5.6$ for NY Vir is larger than 1. We calculated $E_{min}/E_{sec} \approx 1.1$ by using our results ($K_4 = 33.8$ s and $P_4 = 27.2$ yr for the second planet) together with the parameters given in Völschow et al. (2016) for this system. Since the value we obtained is close to 1, we cannot ignore that the additional effect that causes the period change in this system may be due to magnetic activity. In addition, we acknowledge that the mass loss isn't the reason for the orbital period variation in this system because of no mass transfer in PCEBs in which each star is within its Roche lobe (Paczynski 1976; Zorotovic & Schreiber 2013).

Our model predicts that the trend of $O - C$ will be upward and it is an important issue to determine how long this increase will continue. Thus, new observations are needed to constrain the uncertainty of the parameters that will also lead us to understand the structure and nature of NY Vir. For now, even though the two planets model is statistically well consistent with observations, the existence of these planets still remains hypothetical.

## ACKNOWLEDGEMENTS

This work has been supported by The Scientific and Technological Research Council of Turkey (TUBITAK), through project number 114F460 (I.N., H.E.). We thank the team of TUBITAK National Observatory (TUG) for a partial support in using the T100 telescope with project number TUG T100-631 and TUG T100-1333. We also wish to thank Adiyaman University Observatory (Turkey) for the allocation of their telescope time.

*Software*: Python packages (ccdproc (Craig et al. 2017), Astropy (Astropy Collaboration et al. 2013), Numpy (Harris et al. 2020), Photutils (Bradley et al. 2020), LMFIT (Newville et al. 2014), SExtractor (Bertin & Arnouts 1996), emcee (Foreman-Mackey et al. 2013), corner.py (Foreman-Mackey 2016).

## DATA AVAILABILITY

The data underlying this article are available in the article and in its online supplementary material.

This paper has been typeset from a T<sub>E</sub>X/LAT<sub>E</sub>X file prepared by the author.